\documentclass[prl, twocolumn]{revtex4-1}
\usepackage{amsmath,amssymb,bm}
\usepackage{graphicx}
\usepackage{epstopdf}
\usepackage{latexsym}
\usepackage{subfigure}
\usepackage{color}
\usepackage{natbib}
\usepackage{hyperref}
\usepackage{braket}
\usepackage{dsfont}
\hypersetup{
  colorlinks,
  citecolor=magenta,
  linkcolor=blue,
  urlcolor=blue}
      
\newcommand{\cp}[1]{$\mathbb{CP}^{#1}$}

\begin{document}

\title{New easy-plane \cp{N-1} fixed points}
\author{Jonathan D'Emidio}
\affiliation{Department of Physics \& Astronomy, University of
  Kentucky, Lexington, KY-40506-0055}
\author{Ribhu K. Kaul}
\affiliation{Department of Physics \& Astronomy, University of Kentucky, Lexington, KY-40506-0055}
\begin{abstract}
We study fixed points of the easy-plane \cp{N-1}
field theory by combining quantum Monte Carlo
simulations of lattice models of easy-plane
SU($N$) superfluids with field theoretic renormalization group
calculations, by using ideas of deconfined criticality. From our simulations, we present evidence that at small $N$ our lattice model
has a first order
phase transition which progressively weakens as $N$
increases, eventually becoming continuous for large values of $N$.  Renormalization group
calculations in $4-\epsilon$ dimensions provide an explanation of these results as
arising due to the existence of an $N_{ep}$ that separates the fate of the
flows with easy-plane anisotropy. When $N<N_{ep}$ the renormalization group flows
to a discontinuity fixed point and hence a first order
transition arises. On the other hand, for $N > N_{ep}$ the flows are
to a  new easy-plane \cp{N-1} fixed point that describes the quantum
criticality in the lattice model at large $N$.  Our lattice
model at its critical point, thus gives efficient numerical access to a new strongly coupled gauge-matter
field theory.
\end{abstract}
\maketitle

{\em Introduction:} The study of anti-ferromagnets 
has uncovered fascinating connections between quantum spin models and
gauge theories. The connections have allowed novel gauge theoretic
concepts such as deconfinement to be brought into  the realm of condensed
matter physics.
Turning this mapping
around, can the study of
magnetism provide non-perturbative insights into gauge
theories?  Remarkably, advances in
simulation algorithms for quantum anti-ferromagnets~\cite{kaul2013:qmc} have recently
allowed controlled numerical access to otherwise poorly understood strongly coupled gauge theories; the most prominent example being the \cp{N-1} gauge theory proposed
for deconfined critical points (DCP) in SU($N$) magnets~\cite{senthil2004:science}. 

In early work on DCP, a prominent role was played by
the ``easy-plane SU(2)''~\cite{sandvik2002:hjk} magnet and its corresponding ``easy-plane \cp{1}'' field theory~\cite{senthil2004:science,senthil2004:deconf_long}. A self-duality in the field theory suggested that this could be the best
candidate for a deconfined critical point~\cite{motrunich2004:hhog}.  Subsequent numerical work
has concluded however that this transition is first order, both in direct discretizations of the field
theory~\cite{kuklov2006:u1first,kragset2006:first} as well as in
simulations of the quantum
anti-ferromagnet~\cite{demidio2015:epsmN}. The easy-plane case is
in contrast to the symmetric SU($N$) case (we refer to this as s-SU($N$)), where striking agreement
between technical field theoretic calculations~\cite{murthy1990:mono,dyer2015:mono,kaul2008:u1,senthil2006:topo} and numerical
simulations of the quantum magnets has been demonstrated~\cite{kaul2012:j1j2,block2013:fate,nahum2015:so5}. 

The sharp contrast
between the easy-plane and symmetric cases has been unexplained so far.
In this work we address the first order transition in the easy-plane
case using both lattice simulations of an ep-SU($N$) model as well as
renormalization group calculations on a proposed ep-\cp{N-1} field theory.  We find the first order transition
in the ep-SU($N$) models found for $N=2$ in previous work persists for larger $N$. A
careful analysis however shows that the first order jump quantitatively
weakens as $N$ increases. Renormalization group $\epsilon$-expansion
calculations find that the field theory hosts a new ep-\cp{N-1} fixed point only for
$N>N_{\rm ep}$, suggesting that the transition can eventually become continuous. Consistent with this result, we find that the
transition in our lattice model turns continuous around $N\approx 20$. For $N=21$ we
provide a detailed scaling analysis of our numerical data that
confirms a continuous transition in a new universality class. Our work
clarifies and significantly extends the discussion of the DCP phenomena in
easy-plane magnets and its relation to the symmetric case.

{\em Easy-plane model \& field theory:} We consider a family of bipartite
ep-SU($N$) spin models introduced
recently by us~\cite{demidio2015:epsmN}, they are
extensions of the quantum XY model to larger $N$. They are written in terms of the $T^a_i$,
the fundamental generators of SU($N$) on site $i$:
\begin{equation}
\label{eq:j1pj2p}
H =  -\frac{ J_{1\perp}}{N} {\sum_{a,\langle ij\rangle}}^{\prime}  T^a_i   
  T^{a*}_j -\frac{ J_{2\perp}}{N} {\sum_{a,\langle\langle ij\rangle\rangle}}^\prime  T^a_i   
  T^{a}_j.
\end{equation}
the ${\sum}^\prime$ denotes the sum on $a$ is restricted to the $N^2-N$ off-diagonal
generators (a sum on all generators $a$ would give the s-SU($N$)
model). The $\langle ij \rangle$ ($\langle\langle ij\rangle\rangle$)
indicates nearest (next nearest) neighbors on the square lattice which
are on opposite (same) sublattices and in conjugate (same) representations.  The model $H$ is an easy
plane deformation of the s-SU($N$) $J_1$-$J_2$
model~\cite{kaul2012:j1j2}, it has a global
U(1)$^{N-1}\times$S$_N$ in addition to time reversal and lattice symmetries. The model harbors in its phase diagram
the SF-VBS transition for all $N>5$. $H$ is Marshall positive; we
hence simulate it with stochastic series Monte Carlo on $L\times L$
lattice at an inverse temperature $\beta$~\cite{sandvik2010:vietri}.

The effective field theory for SF-VBS phase transition in the
ep-SU($N$) model is obtained  by
applying the ideas of
DCP~\cite{senthil2004:science,senthil2004:deconf_long} to Eq.~(\ref{eq:j1pj2p}). The theory
``ep-\cp{N-1}''  is a sum of kinetic and potential terms
${\cal
  L}_{ep}={\cal L}_1+{\cal L}_2$,
\begin{eqnarray}
\label{eq:cpn-1}
{\cal L}_{1} &=& \sum_\alpha |(\partial_\mu - i e A_\mu)z_\alpha|^2
+ \frac{1}{2} (\vec \nabla \times \vec
A)^2\nonumber\\
 {\cal L}_{2}&=&r\sum_\alpha |z_\alpha|^2+ \frac{u}{2} \left (
\sum_\alpha|z_\alpha|^2\right )^2+\frac{v}{2} \sum_\alpha |z_\alpha|^4,
\end{eqnarray}
where the $z_\alpha$ are $N$ complex fields coupled to a U(1) gauge
field, $A_\mu$. The term $v$ breaks the full s-SU($N$) symmetry of the
s-\cp{N-1} model to a U(1)$^{N-1}\times$S$_N$ (what we shall call ep-SU($N$)). It is known from the
large-$N$ expansion~\cite{halperin1974:largeN} that for $N$ larger
than some finite $N_s$, in $d=3$ the s-\cp{N-1} field theory has a
finite coupling fixed
point (FP). Based on various numerical studies it is now believed that
most likely $N_s<2$,
so that Eq.~\ref{eq:cpn-1} has a FP for all values of
$N$ (see~\cite{nahum2015:deconf} for a nice summary).
A central issue we address here is
the fate of these FPs when easy-plane
anisotropy $v$ is introduced.

{\em Weakening first-order transition:} We begin with a numerical
study of Eq.~(\ref{eq:j1pj2p}).
We have shown in~\cite{demidio2015:epsmN} that the ep-SU($N$) models map to a certain loop model. We can hence
calculate two useful quantities to probe magnetic ordering: the average of the square of the spatial winding number of the loops
$\langle W^2\rangle$ and a normalized magnetic order parameter
$m_\perp^2=\frac{1}{(1-1/N)N^2_{\mathrm{site}}}{\sum_{a}}^{'} \sum_{i,j}
\langle \tilde T^a_i \tilde T^a_j \rangle$ (where the sum on $a$ is on the
off-diagonal generators, $i$ and $j$ are summed on the entire lattice and $\tilde T = T (T^*)$ on the A(B) sublattice), which although off-diagonal in the
$|\alpha\rangle$ basis can be estimated by measuring a particular
statistical property of the loops~\cite{suppmat}. We have normalized $m_\perp^2$ so that the
maximum value it can take is 1 for all $N$, allowing for a
meaningful comparison across different $N$.

\begin{figure}[!t]
\centerline{\includegraphics[angle=0,width=1.0\columnwidth]{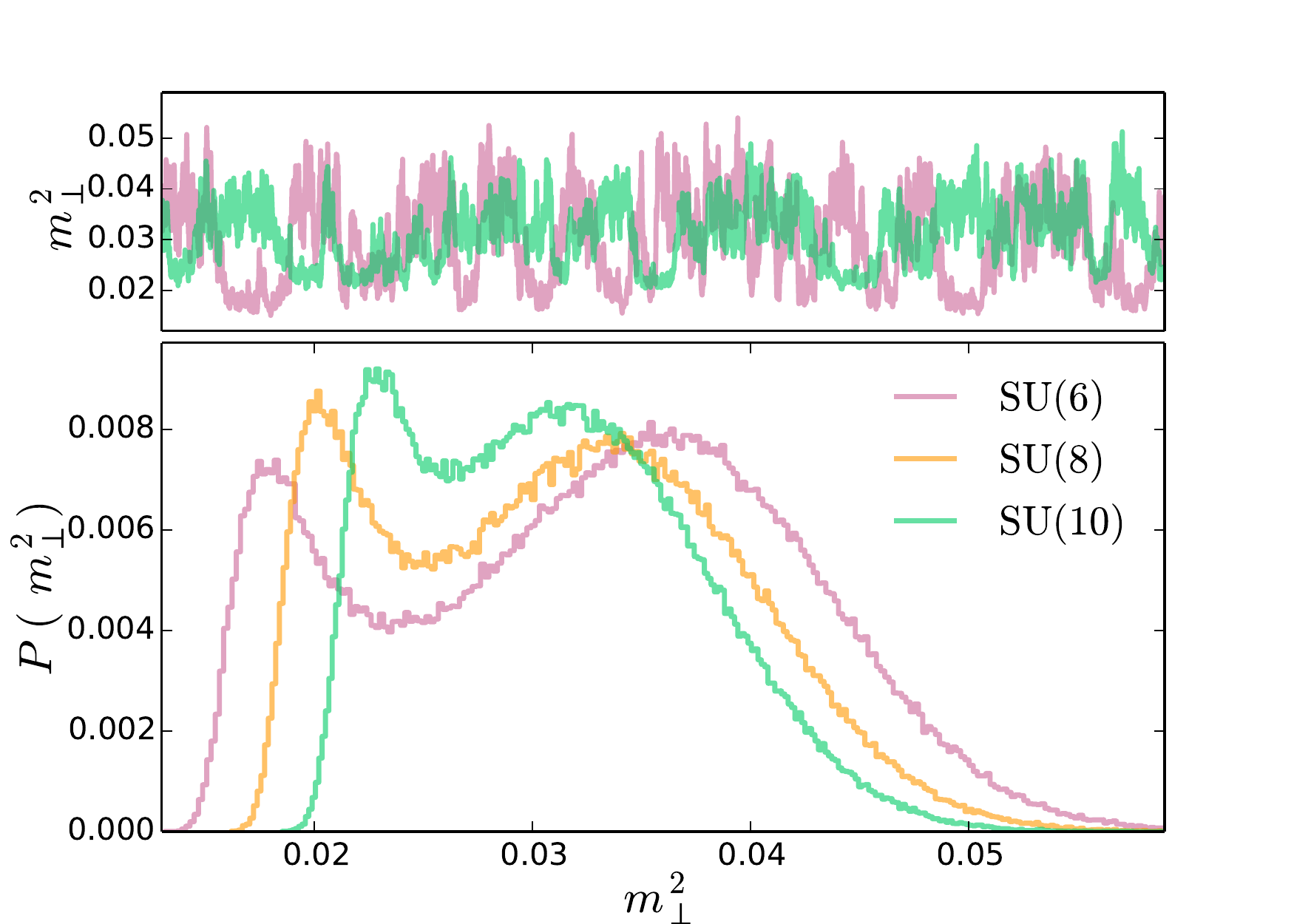}}
\caption{First order transitions for moderate values of $N$. The upper
  panels shows MC histories (arbitrary units) of the estimator for
  $m_\perp^2$ for $N=6$ and 10. The bottom panel shows
  histograms of $m_\perp^2$ taken at $L=50$ for $J_2/J_1\equiv g=0.250,
  0.876, 1.58$ for $N=6,8,10$ respectively, clearly showing
  double peaked behavior.  This data was collected with $\beta=1.5L$. }
\label{fig:hist}
\end{figure}

\begin{figure}[!t]
\centerline{\includegraphics[angle=0,width=1.0\columnwidth]{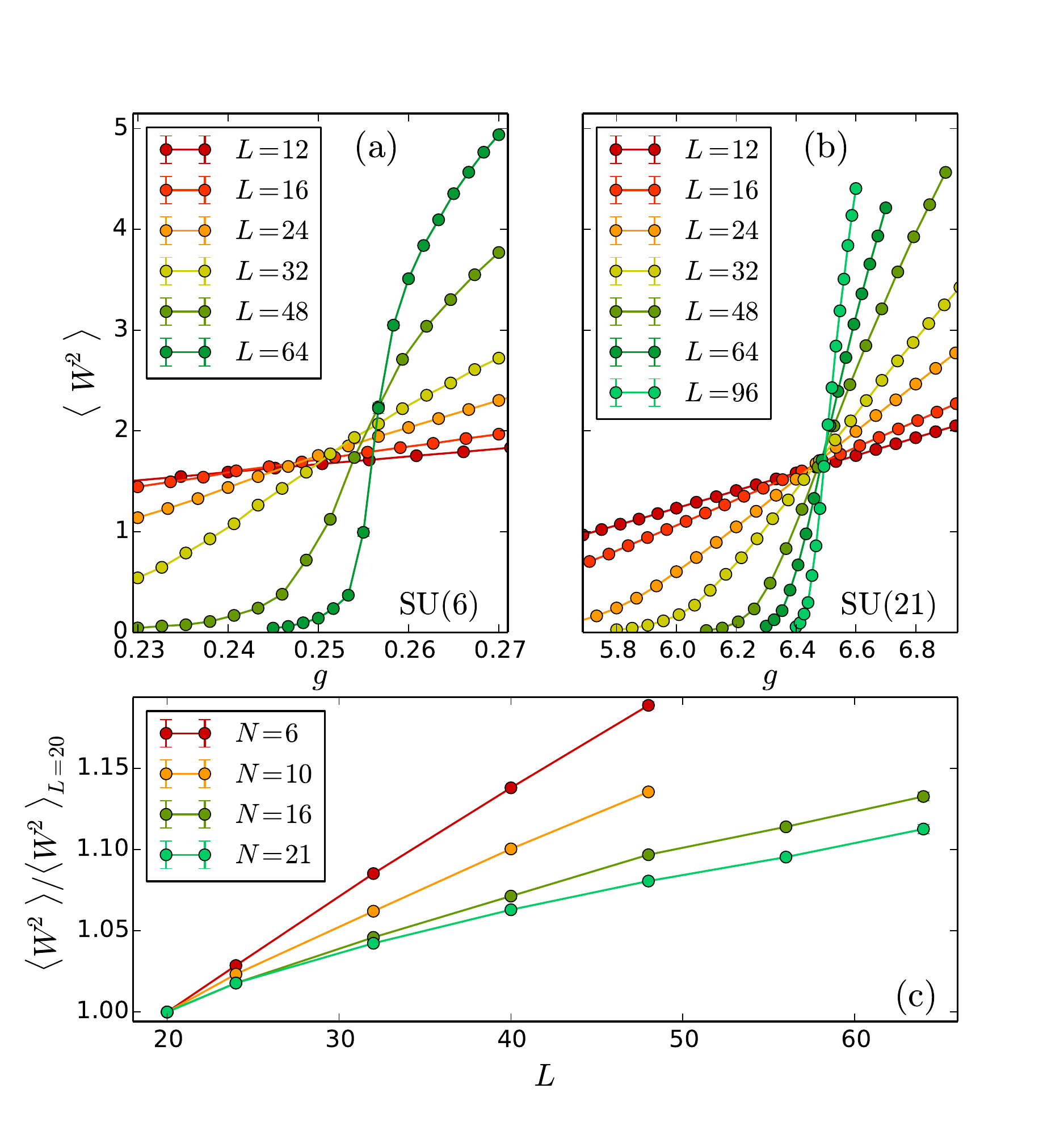}}
\caption{Scaling of the spatial winding number square $\langle
  W^2\rangle$. (a) Crossing for $N=6$. (b) Crossing for
$N=21$. (c) Value at the $L$ and $L/2$ crossing of $\langle
W^2\rangle$ for a range of $N$ normalized to the
crossing value at $L=20$ for each $N$. For the smaller
$N$ a clear linear divergence is seen as expected for a first-order transition (ergodicity issues limit the
system sizes here). For larger $N$ a slow
growth is observed very similar to what has been studied in detail for the
s-SU(2) case and interpreted as evidence for a continuous transition with two
length scales~\cite{Shao2016:two}, like we have here. All the data was taken at $\beta=6L$ which
is in the $T=0$ regime~\cite{suppmat}.  }
\label{fig:Lstiff}
\end{figure}

Previously we found
that the SF-VBS transtion is first order for $N\leq
5$~\cite{demidio2015:epsmN}. In Fig.~\ref{fig:hist} we present data that shows
the first order behavior persists as $N$ is increased up to $N=10$. A hitherto unanswered
but important
question is whether the first order jump weakens as $N$ 
increases. We find evidence in favor of this assertion, since the histogram
peaks get closer as $N$ is increased. Beyond $N\approx 16$ we have found no
evidence for double peaked histograms. To carry out a more quantitative
analysis, which has been popular in the study of the DCPs~\cite{kuklov2006:u1first},
we turn to $\langle W^2\rangle$ (which is
related to the spin stiffness as $\beta\rho_s$). At a
first order transition one expects a linear divergence of $\langle
W^2\rangle$ as one approaches the phase transition since $\rho_s$
stays finite. Any sub-linear behavior indicates that the transition is
continuous since $\rho_s$ vanishes in the thermodynamic limit~\cite{kaul2011:su34}. In Fig.~\ref{fig:Lstiff} we present a study of the crossing
of $\langle
W^2\rangle$. We find clear evidence for the expected linear behavior at
moderate values of $N$. As $N$ is increased beyond about $N\approx 16$ we
find a very slow growth of $\langle
W^2\rangle$ inconsistent with linear behavior but consistent with what has been found in s-SU($N$) models,
where the transition is believed to be
continuous~\cite{kaul2011:su34,Shao2016:two}. This study provides clear evidence
that the first-order jump decreases as $N$ increases, possibly
becoming continuous.

{\em Renormalization group analysis:} The weakening of the first-order SF-VBS
transitions at larger $N$ raises important questions: Is the
transition first order for all $N$ or does it become
continuous beyond some finite $N_{ep}$? If the transition becomes continuous: Is it
truly a new universality class of an ep-\cp{N-1} or does the anisotropy
become irrelevant at the s-FP  resulting in s-\cp{N-1} criticality for the
``easy-plane'' models?

\begin{figure}[!t]
\centerline{\includegraphics[angle=0,width=1.0\columnwidth]{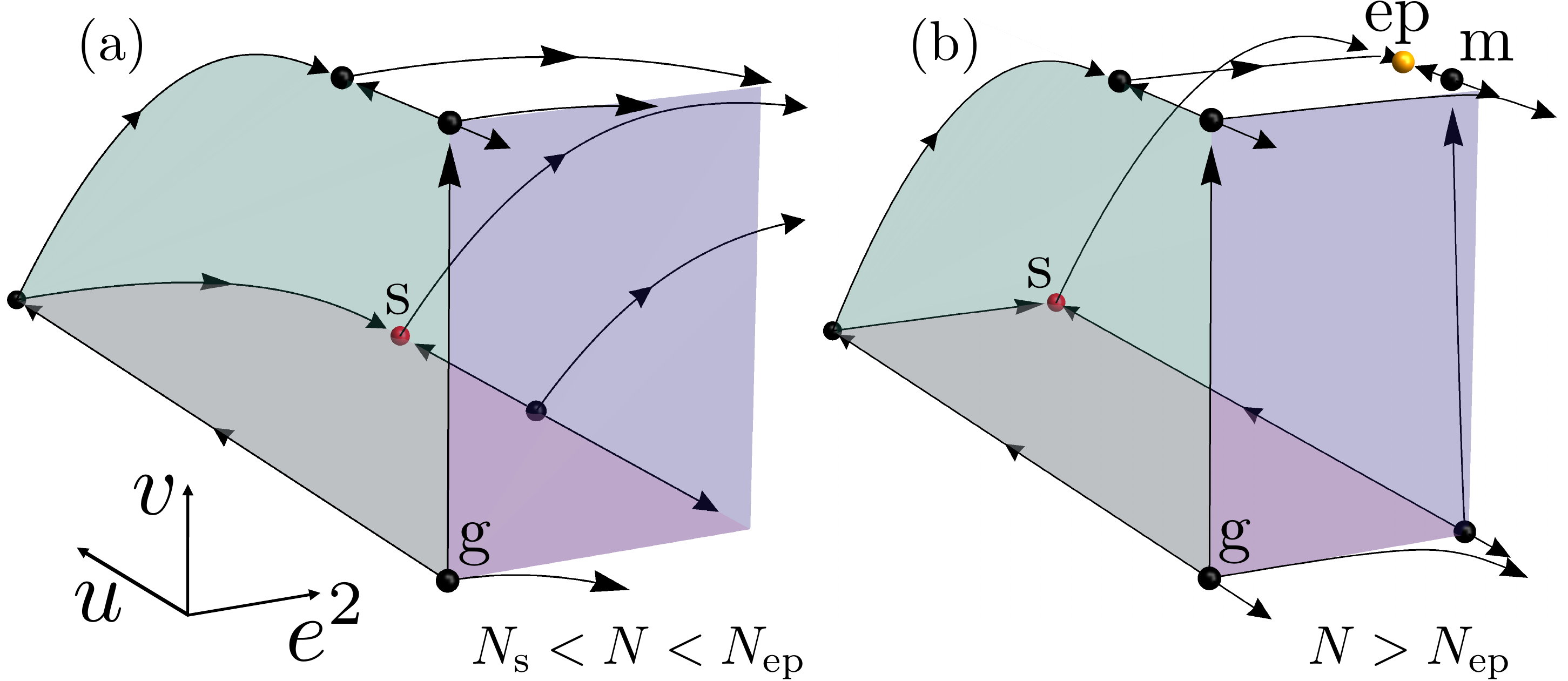}}
\caption{Renormalization group flows of the
  ep-\cp{N-1} model for (a) $N_s<N<N_{ep}$ and (b) $N>N_{ep}$ at
  leading order in $4-\epsilon$ dimensions obtained by numerical
  integration of Eq.(\ref{eq:rg}). Fixed points are shown as bold
  dots, we have only labeled a few significant to our discussion. The flows in the $v=0$ plane have been obtained
  previously~\cite{halperin1974:largeN} and include the ``s'' fixed
  point that describes DCP in s-SU($N$) models (red dot). While the flows have
  many FPs~\cite{suppmat}, a DCP
  of the ep-SU($N$) spin model must have all three eigen-directions in
  the $e^2$-$u$-$v$ irrelevant. For
  $N<N_{ep}$ there are no such FPs; there is hence  a runaway flow to a first order transition. For
  $N>N_{ep}$ two FPs emerge: ``m'' is multicritical and ``ep'' is
the new ep-DCP that describes the SF-VBS transition (yellow dot). The gaussian
fixed point at the origin has been labeled ``g'' for
clarity. More details are in the SM~\cite{suppmat}.}
\label{fig:rgflows}
\end{figure}

To answer these questions, we compute the RG
flows of Eq.~(\ref{eq:cpn-1}) in $4-\epsilon$ dimensions. 
We will work in the critical plane where $r=0$, the $r$ operator being
strongly relevant at tree level will continue to be relevant in the
$\epsilon$-expansion. To leading order (assuming $u$,$v$ and $e^2$
are $\cal{O}(\epsilon)$), we find the following RG equations,
\begin{eqnarray}
\label{eq:rg}
\frac{d e^2}{d {\rm ln} s} &=& \epsilon e^2 - \frac{N}{3}e^4, \nonumber\\
\frac{d u}{d {\rm ln} s} &=& \epsilon u - (N+4)u^2 - 4 uv - 6 e^4 + 6e^2u,
\nonumber\\
\frac{d v}{d {\rm ln} s} &=& \epsilon v - 5v^2 - 6 uv + 6e^2v,
\end{eqnarray}
which for $v=0$ reduce to the well known RG equations for the
s-\cp{N-1} model~\cite{halperin1974:largeN,herbut2010:cp}. Given the relevance of $r$, a generic critical point of ep-\cp{N-1}
would be a fixed
point of Eq.~(\ref{eq:rg}) with all three eigen-directions in
$e^2$-$u$-$v$-space irrelevant. The FP
structure and flows of  Eq.~(\ref{eq:rg}) (shown in Fig.~\ref{fig:rgflows}) change at two values of $N$: $N_s$ and 
$N_{ep}$ with $N_s <N_{ep}$. For $N<N_s$ [not shown] there are no FPs with
$e^2\neq 0$ and a generic flow runs away to a first order
transition. For $N>N_s$ a $v=0$
FP  ``s'' appears, which describes the s-SU($N$) DCP
phenomena, but at which $v$ is always relevant. There are two
distinct fates of the flow with $v\neq 0$: For $N_s<N<N_{ep}$ [see Fig.~\ref{fig:rgflows}(a)] $v$ causes
a runaway flow
to a discontinuity FP, i.e. the phase transition turns first
order. On the other hand, for $N>N_{ep}$ [see
Fig.~\ref{fig:rgflows}(b)] a new fixed point ``ep'' appears. At this
FP all eigen-directions in the $e^2-u-v$ space are irrelevant and
hence $r$ is the only relevant perturbation. ``ep'' hence
describes a generic continuous deconfined SF-VBS transition in models of the form
Eq.~(\ref{eq:j1pj2p}). In the leading order of the
$\epsilon$-expansion we have $N_{s}\approx 183$~\cite{halperin1974:largeN} and $N_{ep}\approx 5363$ (independent of
$\epsilon$). From previous work on the symmetric case, it is well known
that these leading order estimates are unreliable in
$d=3$: Indeed, in the next to leading order, $N_s$
becomes negative for $\epsilon=1$~\cite{kolnberger1990:2lp,herbut1997:epsNs}. Ultimately
the values of $N_{s,ep}$ must be obtained from numerical simulations. 
Nonetheless, it is expected that the basic structure of fixed points
and flows obtained here using the $\epsilon$-expansion are reliable. Based
on our study, we make the following conclusions: Even in a regime where there is a
symmetric fixed point ($N>N_s$), for $N_s<N<N_{ep}$, easy-plane anisotropy will drive the
DCP first-order. For $N>N_{ep}$ a new FP emerges. Easy-plane
anisotropy then results in a continuous SF-VBS transition in a new
ep-\cp{N-1} universality class.

\begin{figure}[!t]
\centerline{\includegraphics[angle=0,width=1.0\columnwidth]{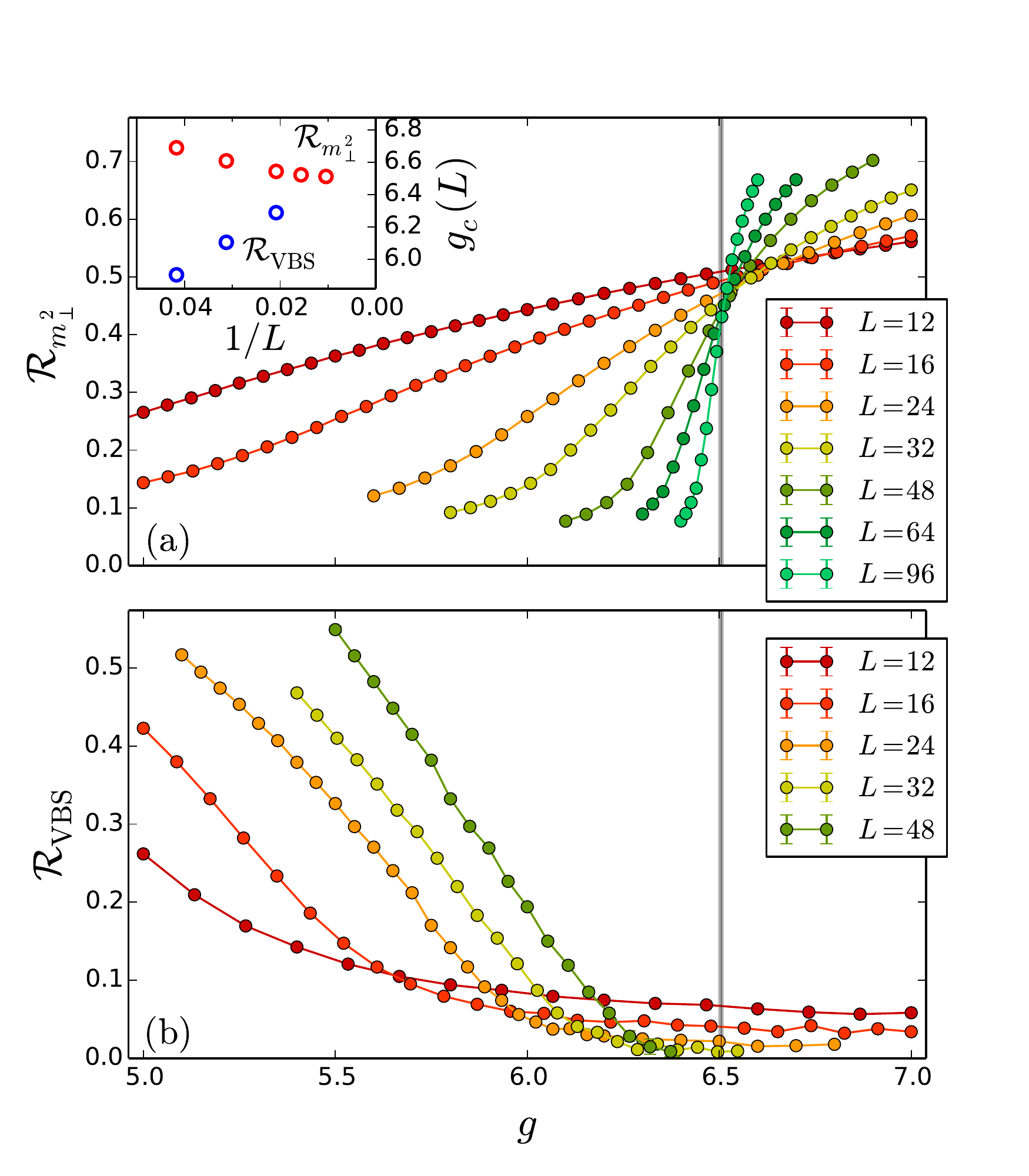}}
\caption{Correlation ratios close to the phase transition for
  $N=21$. (a) The SF order paramater ratio, $R_{m_\perp^2}$ shows good evidence for a continuous
  transition with a nicely convergent crossing point of $g=6.505(5)$. (b) $R_{\rm
    VBS}$ shows a crossing point that converges to the same value of
  the critical coupling. We
  note however that the crossing converges much more slowly (see
  text). The inset shows the convergence of the crossings points of
  $L$ and $L/2$ of SF and
  VBS ratios. Note their convergence to a common critical coupling
  indicating a direct transition.   } 
\label{fig:21cross}
\end{figure}

\begin{figure}[!t]
\centerline{\includegraphics[angle=0,width=1.0\columnwidth]{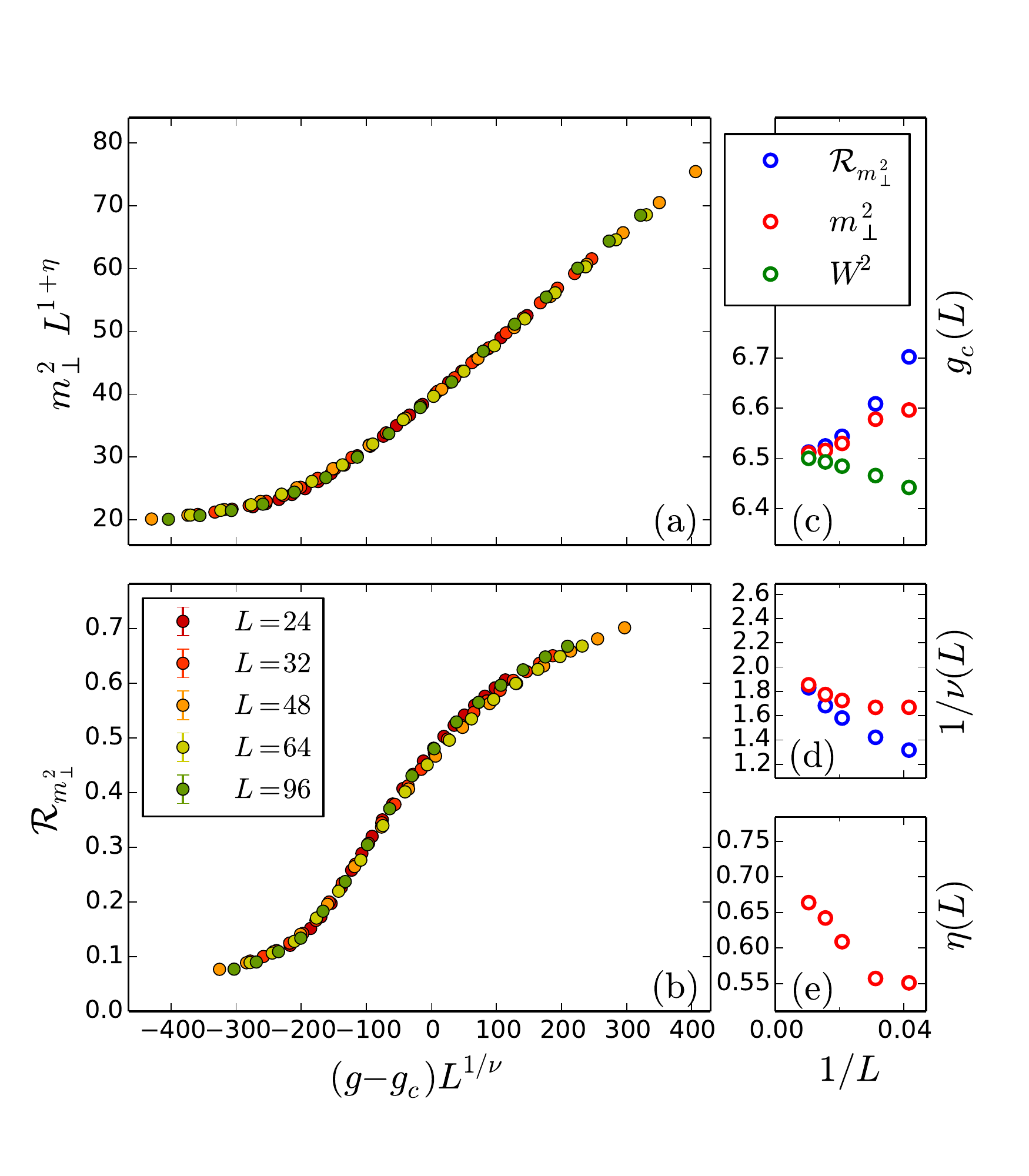}}
\caption{Data collapse for the SF order parameter at $N=21$. (a)
  Finite size data collapsed to 
$m_\perp^2=L^{-(1+\eta_{\rm SF})}
\mathbb{ M} [(g-g_c)L^{1/\nu}]$ with parameters $g_c$=6.511, $\nu$=0.556 (1/$\nu$ = 1.795), $\eta$=0.652. (b) Collapse of ratio $R_{m_\perp^2}=
\mathbb{ R} [(g-g_c)L^{1/\nu}]$ with parameters $g_c$=6.518,
$\nu$=0.582 (1/$\nu$ = 1.719). The side panels shows
convergence of estimates for various quantities from the collapse of
$L$ and $L/2$ data:  (c) the critical coupling $g_{c}=6.505(1)$ from pair-wise
collapses of  $m_\perp^2$ and $R_{m_\perp^2}$, as well as crossings of
$\langle W^2\rangle$ (see Fig.~\ref{fig:Lstiff}) for data. Panel (d) shows $1/\nu(L)$, which we estimate to converge to $1/\nu = 2.3(2)$.  Likewise we estimate $\eta(L)$ to converge to $\eta = 0.72(3)$, which is shown in panel  (e).}
\label{fig:21collapse}
\end{figure}

{\em Study of fixed point:} Having presented evidence from the
$\epsilon$-expansion that with increasing $N$ the transition should turn 
continuous and in a new universality class, it is of interest to study the scaling behavior at  large
$N$. We will focus on $N=21$ where we have found no evidence for first order behavior on the largest
system sizes that we have access to. We
construct dimensionless ratios ${\cal R}_{m_\perp^2}$ and ${\cal R}_{\rm VBS}$
which go to 1(0) in their respective ordered
(disordered) phases. Fig.~\ref{fig:21cross} shows our data for $N=21$. The large
correction to scaling observed in the VBS data are expected: according to
the DCP theory the VBS anomalous dimension $\eta_{\rm VBS}\propto N$ which causes the leading VBS correlation functions
to decay very rapidly at this large value of $N$. This makes it hard to
separate the leading and sub-leading behavior on the available
system sizes. Since the SF data shows a good crossing, we carry
out a full scaling analysis in Fig.~\ref{fig:21collapse}. The
data for both $m_\perp^2$ and $R_{m_\perp^2}$ collapse nicely without
the inclusion of corrections to scaling. They lead to consistent
values of critical couplings and scaling dimensions lending support
for a continuous transition ep-\cp{N-1} fixed point emerging at large
$N$.

In conclusion, we have studied new lattice models for deconfined
criticality with easy-plane
SU($N$) symmetry. We find persistent first order behavior in these lattice
models at small to intermediate $N$, in sharp contrast to the continuous
transitions found in the symmetric models for the same range of $N$. 
As $N$ increases the first order
easy-plane transition weakens and eventually becomes continuous. Our
RG flows provide a way to understand both the first-order and shift to
continuous transitions: The easy-plane anisotropy is
always relevant at the symmetric \cp{N-1} fixed point, for $N<N_{ep}$
there is no easy-plane fixed point and hence the
anisotropy drives the transition first order. For
$N>N_{ep}$ a new fixed point emerges resulting in a continuous
transition in a new ``easy-plane''-\cp{N-1} universality
class which is an example of a strongly coupled gauge-matter field
theory. Our lattice model provides a sign-free discretization of this field
theory that is amenable to efficient numerical simulations. We leave for future work the determination of a
precise value of $N_{ep}$, comparisons of the universal quantities
with easy-plane large-$N$ expansions, and a comparative study of the
scaling corrections between the easy-plane and symmetric
cases. It would be of interest to complement our work with studies of field theories such
as Eq.~(\ref{eq:cpn-1}) using the conformal bootstrap~\cite{elshowk2012:cbs}.

{\em Acknowledgements:}  We thank G. Murthy for many discussions. Partial
financial support was received through NSF DMR-1611161 and the MacAdam
fellowship.  The
numerical simulations reported in the manuscript were carried out on
the DLX cluster at the University of Kentucky.

\bibliography{career}

\clearpage
\widetext

\section{Supplemental materials}

\subsection{Lattice Hamiltonian}

We elaborate on the spin Hamiltonian (\ref{eq:j1pj2p}), which can be simply written in terms of its matrix elements (choosing the normalization $\mathrm{Tr}[T^a T^b] = \delta_{a b}$)

\begin{equation}
\label{eq:j1pj2pmels}
\begin{split}
H =  -\frac{ J_{1\perp}}{N} {\sum_{\langle ij\rangle, \alpha, \beta, \alpha \neq \beta}} |\alpha_i \alpha_j \rangle \langle \beta_i \beta_j| \\
-\frac{ J_{2\perp}}{N} {\sum_{\langle\langle ij\rangle\rangle, \alpha, \beta, \alpha \neq \beta}} |\alpha_i \beta_j \rangle \langle \beta_i \alpha_j|,
\end{split}
\end{equation}

where here we emphasize that $\alpha$ and $\beta$ are summed from 1 to $N$ with the constraint that $\alpha \neq \beta$.  The symmetry of this model is global phase rotations of the form $|\alpha \rangle \to e^{i \theta_{\alpha}} |\alpha \rangle$, where on one sublattice the phase is conjugated due to the representation.  This gives $U(1)^{N-1}$ since an overall phase is trivial.    There is also a discrete permutation symmetry $S_{N}$ that corresponds to a relabeling of the colors.  We note that dropping the constraint $\alpha \neq \beta$ restores the full SU($N$) symmetry and corresponds to a model already studied in the context of deconfined criticality at large $N$ \cite{kaul2012:j1j2}.

This model is explicitly sign free and is amenable to quantum Monte Carlo techniques.  We have used the stochastic series expansion QMC algorithm \cite{syljuasen2002:dirloop}, which samples the partition function at finite temperature.  For practical implementation, one needs to add a constant to the Hamiltonian in order to generate diagonal matrix elements.  We find it convenient to add diagonals with the same weight as the off-diagonals, as follows:

\begin{equation}
\label{eq:j1pj2pmelsShift}
\begin{split}
H \to  -\frac{ J_{1\perp}}{N} \sum_{\langle ij\rangle} \left(\sum_{\alpha, \beta, \alpha \neq \beta} |\alpha_i \alpha_j \rangle \langle \beta_i \beta_j| + \mathds{1} \right)\\
-\frac{ J_{2\perp}}{N} \sum_{\langle\langle ij\rangle\rangle} \left(\sum_{\alpha, \beta, \alpha \neq \beta} |\alpha_i \beta_j \rangle \langle \beta_i \alpha_j| + \mathds{1} \right).
\end{split}
\end{equation}

We refer the reader to more details of the loop algorithm contained in \cite{demidio2015:epsmN}.  One notable aspect of this $J_{1\perp}-J_{2 \perp}$ model is that the addition of the $J_{2 \perp}$ term can be treated with minimal extra effort, given a code that simulates $J_{1\perp}$ only.  Updating a matrix element (vertex) associated with $J_{2 \perp}$ can be achieved by first time reversing the spin states on one sublattice of the vertex, then scattering through the vertex according to the rules for $J_{1 \perp}$ matrix elements, and finally reversing the spins back.  

\subsection{Measurements}

Many of our measurements are part of the standard tool kit.  This includes the winding number fluctuation $\langle W^2 \rangle$, which is related to the superfluid stiffness $\rho=\langle W^2 \rangle / \beta$; and also the equal-time bond-bond correlation function, which is used to construct $\mathcal{R}_{\mathrm{VBS}}$.  In order to introduce the VBS order parameter and ratio, we first consider the Fourier transformed bond-bond correlator 
{\allowdisplaybreaks
\begin{equation}
		\tilde{C}^{a}_{VBS}(\vec{q})=\frac{1}{N_{\mathrm{site}}^2}\sum_{\vec{r},\vec{r}'}e^{i(\vec{r}-\vec{r}')\cdot\vec{q}}\langle
                \tilde{P}_{\vec{r} a}\tilde{P}_{\vec{r}' a}\rangle,
\label{eq:FT}
\end{equation}}

where $\tilde{P}_{\vec{r} a}$ is an off-diagonal nearest neighbor bond operator of the form $\sum_{\alpha, \beta,\alpha \neq \beta}^N |\alpha \alpha\rangle_{ij}\langle \beta \beta |_{ij}$ that acts at a bond location $\vec{r}$ with orientation $a \in \{x,y\}$.

For columnar VBS patterns, peaks appear at the momenta $(\pi,0)$ and $(0,\pi)$ for $x$ and $y$-oriented bonds, respectively.  The VBS order parameter is thus given by
{\allowdisplaybreaks
\begin{equation}
		\mathcal{O}_{VBS}=\frac{\tilde{C}^{x}_{VBS}(\pi,0)+\tilde{C}^{y}_{VBS}(0,\pi)}{2}.
\label{eq:Ovbs}
\end{equation}}

We can further construct the VBS ratio, which is defined as
{\allowdisplaybreaks
\begin{equation}
		\mathcal{R}^x_{VBS}=1- \tilde{C}^{x}_{VBS}(\pi+2\pi/L,0)/\tilde{C}_{VBS}^{x}(\pi,0)
\label{eq:RXvbs}
\end{equation}}
And similarly for $\mathcal{R}^y_{VBS}$ with all of the $q_x$ and $q_y$ arguments swapped.  We then average over $x$ and $y$- orientations.
{\allowdisplaybreaks
\begin{equation}
		\mathcal{R}_{VBS}=\frac{\mathcal{R}^x_{VBS}+\mathcal{R}^y_{VBS}}{2}.
\label{eq:RXvbs}
\end{equation}}
This quantity goes to 1 in a phase with long-range VBS order, and
approaches 0 in the superfluid phase.  It is thus a useful crossing quantity that allows us to locate the transition.

To construct these quantities, we measure the equal time bond-bond correlation function in QMC with the following estimator
{\allowdisplaybreaks
\begin{equation}
		\langle \Theta_1 \Theta_2\rangle=\frac{1}{\beta^2} \langle (n-1)! N[\Theta_1,\Theta_2] \rangle
\label{eq:QMCcorr}
\end{equation}}
where $\Theta_1$ and $\Theta_2$ are any two QMC operators (in our case off-diagonal nearest neighbor bond operators), $n$ is the number of non-null operators in the operator string, and $N[\Theta_1,\Theta_2]$ is the number of times $\Theta_1$ and $\Theta_2$ appear in sequence in the operator string (excluding null slots).

\begin{figure}[!t]
\centerline{\includegraphics[angle=0,width=0.6\columnwidth]{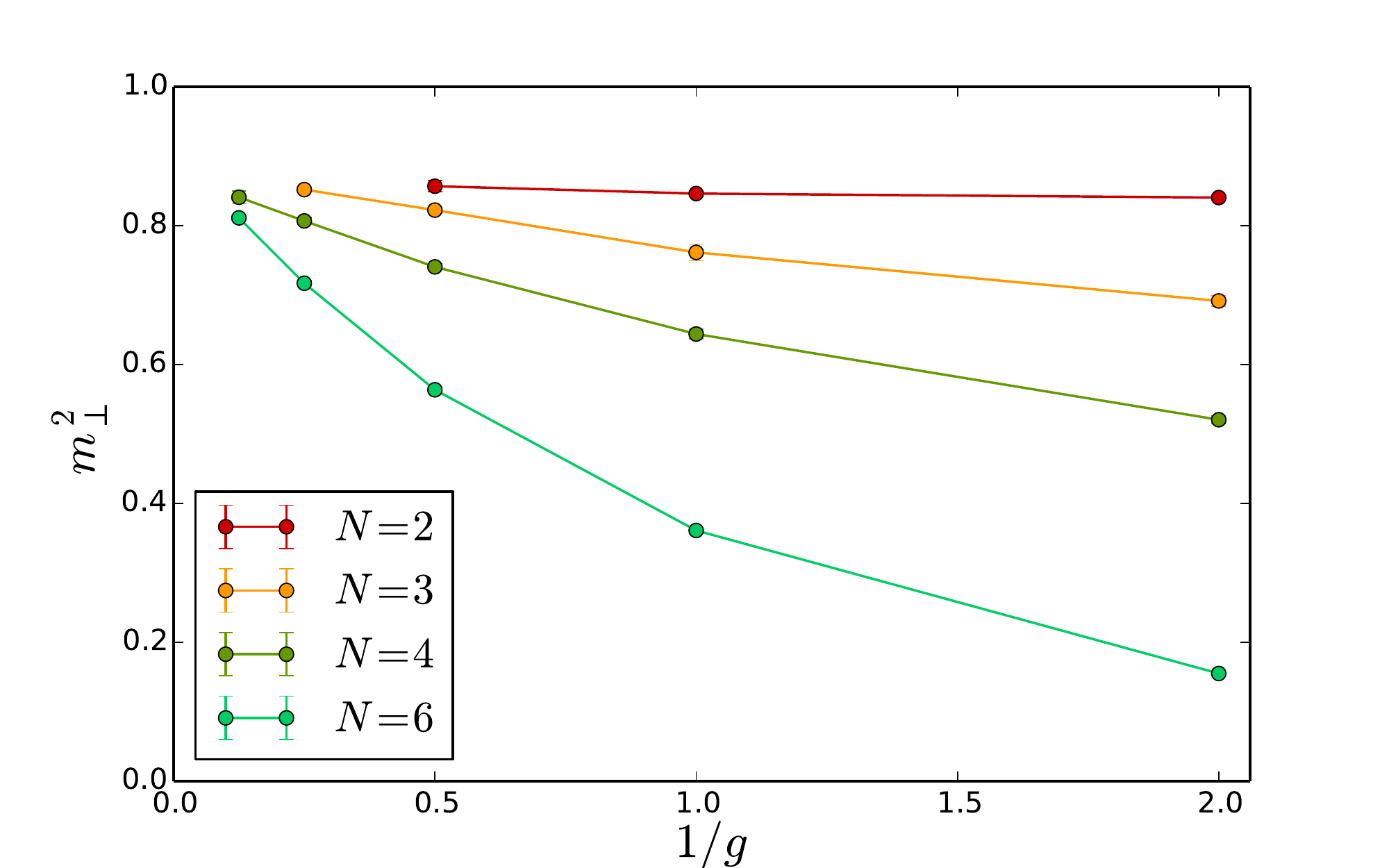}}
\caption{Maximum values of $m^2_{\perp}$ deep in the superfluid phase for different values of $N$.  Each data point was obtained by extrapolating the value of $m^2_{\perp}$ in the thermodynamic limit at a fixed value of the coupling $g=J_{2\perp}/J_{1\perp}$.  Increasing $g$ drives the system into the superfluid phase, we can thus observe the maximum possible value of $m^2_{\perp}$ in the limit $1/g \to 0$.  The data is consistent with an upper bound of one.}
\label{fig:msqextrap}
\end{figure} 

\begin{table}
\centering
\begin{tabular}{||l l l l l||}
\noalign{\hrule height 2pt}
\multicolumn{1}{||c|}{$4 \times 4$} & \multicolumn{1}{c|}{$N=2$} & \multicolumn{1}{c|}{$J_{1\perp}=1.0$} & \multicolumn{1}{c||}{$J_{2\perp}=1.0$}\\
\noalign{\hrule height 2pt}
\multicolumn{1}{||c|}{$e_{\mathrm{ex}}$} & \multicolumn{1}{l|}{$-1.082912818$} & \multicolumn{1}{c|}{$m^2_{\perp\mathrm{ex}}$} & \multicolumn{1}{l||}{$1.110134109$} \\
\multicolumn{1}{||c|}{$e_{\mathrm{QMC}}$} & \multicolumn{1}{l|}{$-1.082909(4)$} & \multicolumn{1}{c|}{$m^2_{\perp\mathrm{QMC}}$} & \multicolumn{1}{l||}{$1.11014(1)$}\\
\noalign{\hrule height 2pt}
\multicolumn{1}{||c|}{$4 \times 4$} & \multicolumn{1}{c|}{$N=2$} & \multicolumn{1}{c|}{$J_{1\perp}=1.0$} & \multicolumn{1}{c||}{$J_{2\perp}=2.0$}\\
\noalign{\hrule height 2pt}
\multicolumn{1}{||c|}{$e_{\mathrm{ex}}$} & \multicolumn{1}{l|}{$-1.629091615$} & \multicolumn{1}{c|}{$m^2_{\perp\mathrm{ex}}$} & \multicolumn{1}{l||}{$1.107517598$}\\
\multicolumn{1}{||c|}{$e_{\mathrm{QMC}}$} & \multicolumn{1}{l|}{$-1.629086(5)$} & \multicolumn{1}{c|}{$m^2_{\perp\mathrm{QMC}}$} & \multicolumn{1}{l||}{$1.10753(1)$}\\
\noalign{\hrule height 2pt}
\multicolumn{1}{||c|}{$4 \times 2$} & \multicolumn{1}{c|}{$N=3$} & \multicolumn{1}{c|}{$J_{1\perp}=1.0$} & \multicolumn{1}{c||}{$J_{2\perp}=1.0$}\\
\noalign{\hrule height 2pt}
\multicolumn{1}{||c|}{$e_{\mathrm{ex}}$} & \multicolumn{1}{l|}{$-1.131110222$} & \multicolumn{1}{c|}{$m^2_{\perp\mathrm{ex}}$} & \multicolumn{1}{l||}{$1.408846135$}\\
\multicolumn{1}{||c|}{$e_{\mathrm{QMC}}$} & \multicolumn{1}{l|}{$-1.131100(5)$} & \multicolumn{1}{c|}{$m^2_{\perp\mathrm{QMC}}$} & \multicolumn{1}{l||}{$1.408849(6)$} \\
\noalign{\hrule height 2pt}
\multicolumn{1}{||c|}{$4 \times 2$} & \multicolumn{1}{c|}{$N=3$} & \multicolumn{1}{c|}{$J_{1\perp}=1.0$} & \multicolumn{1}{c||}{$J_{2\perp}=2.0$}\\
\noalign{\hrule height 2pt}
\multicolumn{1}{||c|}{$e_{\mathrm{ex}}$} & \multicolumn{1}{l|}{$-1.618465034$} & \multicolumn{1}{c|}{$m^2_{\perp\mathrm{ex}}$} & \multicolumn{1}{l||}{$1.423156633$}\\
\multicolumn{1}{||c|}{$e_{\mathrm{QMC}}$} & \multicolumn{1}{l|}{$-1.618476(7)$} & \multicolumn{1}{c|}{$m^2_{\perp\mathrm{QMC}}$} & \multicolumn{1}{l||}{$1.423150(8)$} \\
\hline
\end{tabular}
\caption{QMC versus exact diagonalization.  For brevity we provide just the energy per site and normalized $m^2_{\perp}$ for $N=2$ and $N=3$ systems.  We have used $\beta=32$ in our QMC simulations.}
\label{tab:ed}
\end{table}

\begin{figure}[!t]
\centerline{\includegraphics[angle=0,width=0.6\columnwidth]{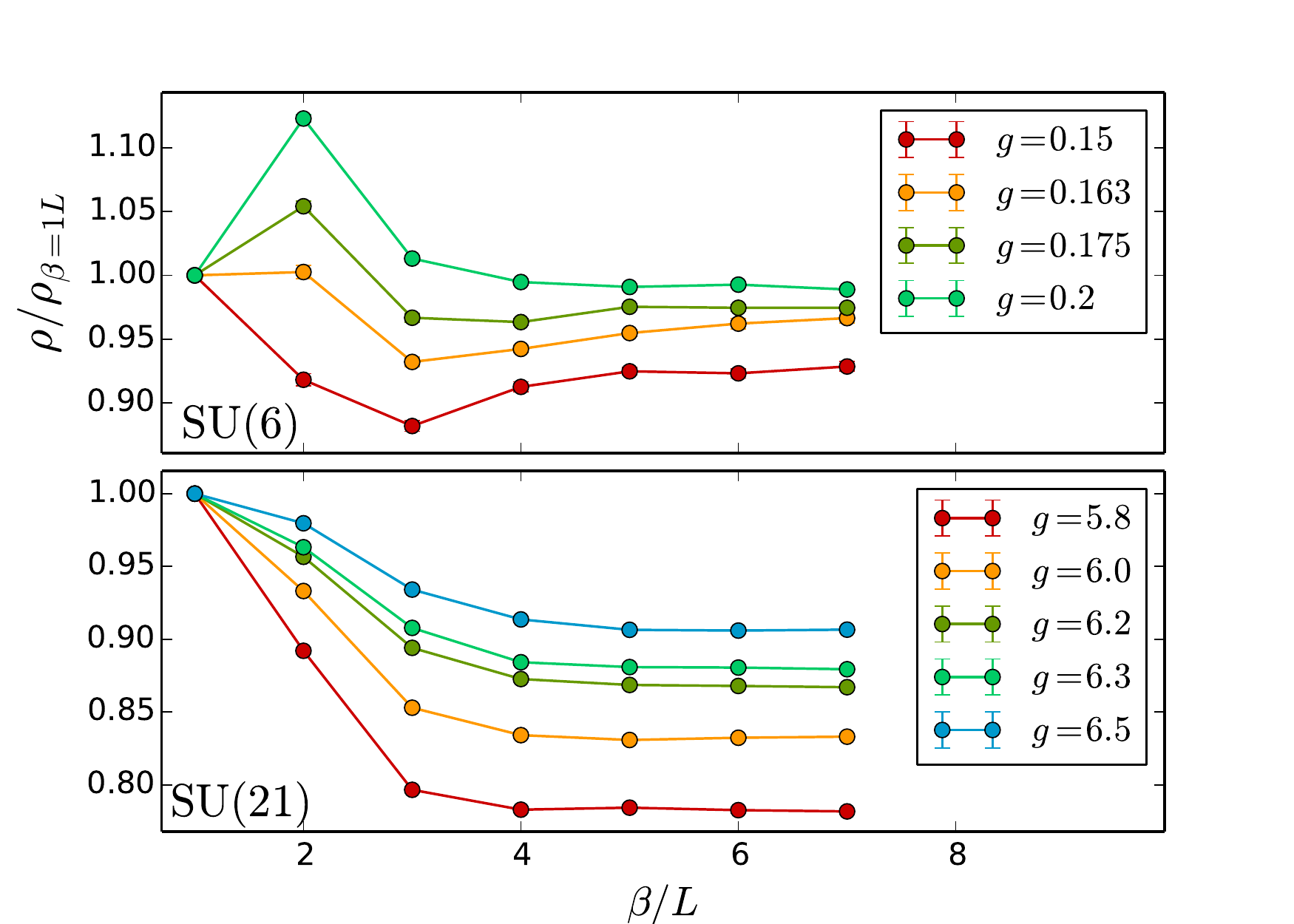}}
\caption{Convergence of the superfluid stiffness ($\rho$) as a function of $\beta$ for $L=16$ around transition.  Finite temperature effects are absent when $\beta \approx 5L$ for both SU(6) and SU(21).  We therefore conservatively fix $\beta=6L$ for the crossing analysis and data collapse presented in the main paper. }
\label{fig:stiffvsbeta}
\end{figure}

In this work we have also made use of the less common``in-plane" magnetization $m^2_{\perp}$, for which we have produced magnetic data collapses.  We now outline this particular measurement.  The reader is directed to \cite{dorneich2001:Dynamics} for more details.

The superfluid (magnetic) ordering in our system is off-diagonal in the computational basis, meaning that the relevant equal-time correlation functions are of the form $\langle S^{+}_i S^{-}_j \rangle$, written in terms of raising and lowering spin operators.  The placement of such an operator into a QMC configuration will in general give zero, unless the two operators are joined by a loop of a certain color.  Loop updates can be regarded as inserting a raising and lowering pair (defects) at the same time slice and spatial location, then propagating one of the defects until it annihilates with the first, forming a closed loop.  The algorithm is stochastically sampling the space of allowed configurations with two defects.  The measurement $\langle S^{+}_i S^{-}_j \rangle$ is then given by the average number of times these two defects occur at the same time slice at locations $i$ and $j$, which is averaged over the total number of loops grown.  We note that starting loops at a vertex leg will bias this measurement, an so it must be performed by choosing a random time slice and spatial location in which to start the loop.

We normalize the (0,0) component of the correlation function to the value $N-1 = \sum_{a}^{\prime} T^a T^a$, where the sum on $a$ is on the off-diagonal generators.  This is due to the fact that we have chosen $\mathrm{Tr}[T^a T^b] = \delta_{a b}$.  We then construct $m^2_{\perp}$ based on the off-diagonal correlation function as follows:

\begin{equation}
\label{eq:msq}
m_\perp^2=\frac{1}{(1-\frac{1}{N})N^2_{\mathrm{site}}}{\sum_{a}}^{'} \sum_{i,j} \langle \tilde T^a_i \tilde T^a_j \rangle
\end{equation}

where $i$ and $j$ are summed on the entire lattice and $\tilde T = T (T^*)$ on the A(B) sublattice.  In practice we use lattice symmetries to reduce the number of correlators that need to be stored.  Here we have importantly chosen an overall normalization factor for $m_\perp^2$ such that the upper bound in the ordered phase is equal to one for all $N$.  The factor $(1-1/N)=(N-1)/N$ accounts for the fact that the generators have been normalized such that the (0,0) component of the correlator is $N-1$ and that, given a color at location 0, the probability of picking the same color at long distances is $1/N$.  This amounts to saying that the most magnetically ordered configurations consist of only $N$ loops (one for each color).

The normalization allows us to meaningfully compare this measurement across different values of $N$ and observe a reduction in the size of the first-order jump as in Fig. \ref{fig:hist}.  We demonstrate that this normalization is correct in Fig. \ref{fig:msqextrap} by extrapolating the value of $m_\perp^2$ in the thermodynamic limit for different values of $g= J_{2\perp}/J_{1\perp}$.

Analogous to the VBS ratio, we can construct the magnetic ratio as well ($\mathcal{R}_{m^2_{\perp}}$).  If we denote the Fourier transformed off-diagonal spin-spin correlator as

\begin{equation}
		\tilde{C}_{m^2_{\perp}}(\vec{q})=\frac{1}{(1-\frac{1}{N})N_{\mathrm{site}}^2} {\sum_{a}}^{'} \sum_{\vec{r},\vec{r}'}e^{i(\vec{r}-\vec{r}')\cdot\vec{q}}\langle  \tilde T^a_{\vec{r}} \tilde T^a_{\vec{r}'}  \rangle,
\label{eq:FTmsq}
\end{equation}

then the magnetic ratio is then given by

{\allowdisplaybreaks
\begin{equation}
		\mathcal{R}_{m^2_{\perp}}=1- \frac{\tilde{C}_{m^2_{\perp}}(2\pi/L,0) + \tilde{C}_{m^2_{\perp}}(0,2\pi/L)}{2\tilde{C}_{m^2_{\perp}}(0,0)}.
\label{eq:Rmsq}
\end{equation}}

We have thoroughly checked all of our measurements against exact diagonalization on small system sizes.  In Table \ref{tab:ed} we provide comparisions of the energy per site and normalized $m^2_{\perp}$ between QMC and exact diagonalization, showing agreement within the statistical error.

In order to avoid finite temperature effects in our crossing analysis and data collapse, we fixed $\beta =6L$ in those simulations.  This value was chosen based on the zero temperature convergence of the superfluid stiffness $\rho=\langle W^2 \rangle / \beta$ on an $L=16$ system size for $N=6$ and $N=21$ around the transition.  This data is shown in Fig. \ref{fig:stiffvsbeta}.

\subsection{Renormalization group methods}

We refer the reader to \cite{herbut2010:cp}, which is a useful reference that outlines in detail many of the results that we will now discuss.  We have performed momentum shell renormalization group transformations in $4-\epsilon$ dimensions with the Lagrangian density (\ref{eq:cpn-1}).  We therefore as a starting point write the Euclidean action in $k-$space as follows:
    
\begin{equation}
\label{eq:kaction}
\begin{split}
S=&\sum_{\alpha}\int \frac{\mathrm{d}\vec{k}}{(2 \pi)^d}(\vec{k}^2+r)z^{*}_{\alpha}(\vec{k}) z_{\alpha}(\vec{k})\\
- e&\sum_{\alpha}\int \frac{\mathrm{d}\vec{k} \, \mathrm{d}\vec{p}}{(2 \pi)^{2d}}(2\vec{k}+\vec{p}) \cdot \vec{A}(\vec{p})z^{*}_{\alpha}(\vec{p}+\vec{k}) z_{\alpha}(\vec{k})\\
+ e^2 &\sum_{\alpha}\int \frac{\mathrm{d}\vec{k} \, \mathrm{d}\vec{p} \, \mathrm{d}\vec{q}}{(2 \pi)^{3d}} \vec{A}(\vec{p}) \cdot \vec{A}(\vec{q})z^{*}_{\alpha}(\vec{p}+\vec{q}+\vec{k}) z_{\alpha}(\vec{k})\\
+ \frac{1}{2} &\sum_{\alpha, \beta} (u + v \delta_{\alpha \beta})\int \frac{\mathrm{d}\vec{k}_1 \ldots \mathrm{d}\vec{k}_4}{(2 \pi)^{3d}} \delta^d (\vec{k}_1-\vec{k}_2+\vec{k}_3-\vec{k}_4)z^{*}_{\alpha}(\vec{k}_1) z_{\alpha}(\vec{k}_2) z^{*}_{\beta}(\vec{k}_3) z_{\beta}(\vec{k}_4) \\
+\frac{1}{2} &\sum_{i,j} \int \frac{\mathrm{d}\vec{k}}{(2 \pi)^d} A_i(\vec{k})[\vec{k}^2 (\delta_{ij}-\hat{k}_i \hat{k}_j)+\frac{k_i k_j}{\xi}]A_j(-\vec{k}).
\end{split}
\end{equation}

\begin{figure}[!t]
\centerline{\includegraphics[angle=0,width=0.6\columnwidth]{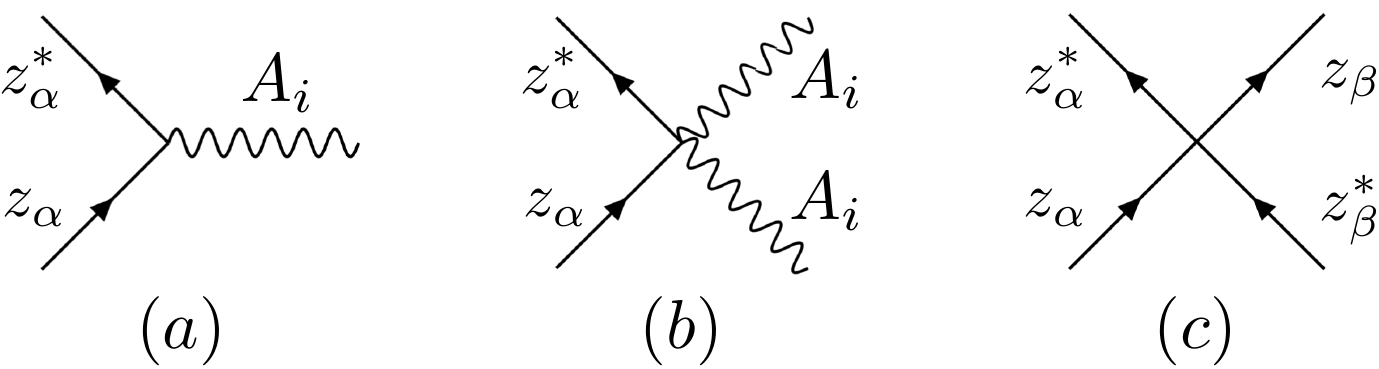}}
\caption{Interaction vertices that couple fast and slow modes.  Perturbative RG is carried out to first order in $\epsilon$ by forming all possible one loop diagrams from these vertices.}
\label{fig:RGvertices}
\end{figure}

Here we have used the standard Faddeev-Popov gauge fixing trick, with gauge fixing parameter $\xi$ \cite{peskin1995:QFT}.  Fig. \ref{fig:RGvertices} shows the interaction vertices that couple the slow and fast Fourier modes, where the slow modes (to be denoted schematically by $z_{<}$ and $A_{<}$)  have momenta in the range $0<|\vec{k}| < \Lambda / s$ and the fast modes ($z_{>}$ and $A_{>}$) have momenta $\Lambda / s <|\vec{k}|<\Lambda$.

The partition function is separated according to fast and slow modes and the contribution from the fast modes is evaluated perturbatively assuming $u$, $v$ and $e^2$ are all of order $\epsilon$.

\begin{equation}
\label{eq:Zfastslow}
Z=\int\mathcal{D}z^{*}_{<}\mathcal{D}z_{<}\mathcal{D}A_{<} e^{-S_{<}}\int\mathcal{D}z^{*}_{>}\mathcal{D}z_{>}\mathcal{D}A_{>} e^{-(V_{<,>} + V_{>,>})}  e^{-S_{0,>}}.
\end{equation}

We have separated out the part of the action which only depends on the slow modes ($S_{<}$), as well as interaction terms in $S$ which mix slow and fast ($V_{<,>}$) and interactions for the fast modes ($V_{>,>}$).  Additionally, the part of the action that is quadratic in the fast modes ($S_{0,>}$) is explicitly seperated out.  The average ${\langle e^{-(V_{<,>} + V_{>,>})} \rangle}_{0,>}$ can then be computed perturbatively using the cumulant expansion.  This leads to a renormalization of the terms in $S_{<}$.  To first order in $\epsilon$ the renormalized action for the slow modes ($S^{\prime}_{<}$) looks like

\begin{equation}
\label{eq:slowkaction}
\begin{split}
S^{\prime}_{<}=&\sum_{\alpha}\int^{\Lambda/ s}_0 \frac{\mathrm{d}\vec{k}}{(2 \pi)^d}(\mathcal{Z}_{\eta}\vec{k}^2+ \mathcal{Z}_{r}r)z^{*}_{\alpha}(\vec{k}) z_{\alpha}(\vec{k})\\
- \mathcal{Z}_{\eta}e&\sum_{\alpha}\int^{\Lambda/ s}_0 \frac{\mathrm{d}\vec{k} \, \mathrm{d}\vec{p}}{(2 \pi)^{2d}}(2\vec{k}+\vec{p}) \cdot \vec{A}(\vec{p})z^{*}_{\alpha}(\vec{p}+\vec{k}) z_{\alpha}(\vec{k})\\
+ \mathcal{Z}_{\eta}e^2 &\sum_{\alpha}\int^{\Lambda/ s}_0 \frac{\mathrm{d}\vec{k} \, \mathrm{d}\vec{p} \, \mathrm{d}\vec{q}}{(2 \pi)^{3d}} \vec{A}(\vec{p}) \cdot \vec{A}(\vec{q})z^{*}_{\alpha}(\vec{p}+\vec{q}+\vec{k}) z_{\alpha}(\vec{k})\\
+ \frac{1}{2} &\sum_{\alpha, \beta} (u^{\prime} + v^{\prime} \delta_{\alpha \beta})\int^{\Lambda /s}_0 \frac{\mathrm{d}\vec{k}_1 \ldots \mathrm{d}\vec{k}_4}{(2 \pi)^{3d}} \delta^d (\vec{k}_1-\vec{k}_2+\vec{k}_3-\vec{k}_4)z^{*}_{\alpha}(\vec{k}_1) z_{\alpha}(\vec{k}_2) z^{*}_{\beta}(\vec{k}_3) z_{\beta}(\vec{k}_4) \\
+\frac{1}{2} &\sum_{i,j} \int^{\Lambda /s}_{0} \frac{\mathrm{d}\vec{k}}{(2 \pi)^d} A_i(\vec{k})[\mathcal{Z}_{A} \vec{k}^2 (\delta_{ij}-\hat{k}_i \hat{k}_j)+\frac{k_i k_j}{\xi}]A_j(-\vec{k}),
\end{split}
\end{equation}

where

\begin{equation}
\label{eq:Zfactors}
\begin{split}
\mathcal{Z}_{\eta}&=1-3\hat{e}^2 \text{ln}(s)\\
\mathcal{Z}_{r}&=1-[(N+1)\hat{u}+2\hat{v}] \text{ln}(s)\\
\mathcal{Z}_{A}&=1+\frac{1}{3} N \hat{e}^2 \text{ln}(s)\\
\hat{u}^{\prime}&=\hat{u}-[(N+4)\hat{u}^2 +4 \hat{u} \hat{v} +6 \hat{e}^4] \text{ln}(s)\\
\hat{v}^{\prime}&=\hat{v}-[5\hat{v}^2 + 6\hat{u}\hat{v}]\text{ln}(s).
\end{split}
\end{equation}

Here we have defined dimensionless couplings $\hat{g} \equiv g \Lambda^{d-4}\mathcal{S}_d/(2\pi)^d$ with $\mathcal{S}_d$ being the surface area of a $d-$dimensional sphere and $g=e^2,u,v$.  We have used Landau gauge $\xi=0$ (which forces $\vec{\nabla} \cdot \vec{A} = 0$) in order to evaluate loop integrals involving the gauge field propagator.  We now rescale the momentum: $\vec{k} \to \vec{k}/s$ and fields $z_{\alpha}(\vec{k}/s) \to s^{(d/2 +1)}\mathcal{Z}^{-1/2}_{\eta} z_{\alpha}(\vec{k})$, $\vec{A}(\vec{k}/s) \to s^{(d/2 +1)}\mathcal{Z}^{-1/2}_{A} \vec{A}(\vec{k})$.  This defines the renormalized couplings as $r(s)=\mathcal{Z}_{r} s^2 r / \mathcal{Z}_{\eta}$, $u(s)= u^{\prime} s^{\epsilon}/\mathcal{Z}^2_{\eta}$, $v(s)= v^{\prime} s^{\epsilon}/\mathcal{Z}^2_{\eta}$ and $e^2(s) = e^2 s^{\epsilon}/\mathcal{Z}_{A}$.  With the renormalized couplings we can now write the $\beta -$ functions:

\begin{subequations}
\begin{align}
\frac{d r}{d {\rm ln} s} &= r [2 - (N+1)u -2v +3e^2] \label{eq:RGfull1} \\
\frac{d e^2}{d {\rm ln} s} &= \epsilon e^2 - \frac{N}{3}e^4 \label{eq:RGfull2} \\
\frac{d u}{d {\rm ln} s} &= \epsilon u - (N+4)u^2 - 4 uv - 6 e^4 + 6e^2u \label{eq:RGfull3} \\
\frac{d v}{d {\rm ln} s} &= \epsilon v - 5v^2 - 6 uv + 6e^2v. \label{eq:RGfull4}
\end{align}
\end{subequations}

We note that when $N=1$, these equations have the the same form for $v$ as for $u$.  This is a useful check, as the $v$ term is identical to the $u$ term when $N=1$.  A drawback of the momentum shell approach is that the momentum cutoff breaks gauge invariance, and hence we have discarded terms that would renormalize $\xi$.  The field theoretic RG formulation preserves gauge invariance, and we have checked that out $\beta - $ functions match using this approach as well.

\subsection{RG flow equations}

Here we will elaborate on the structure of the RG flow equations that have been presented in the main text.  Throughout this discussion, we will take $r=0$, working in the critical plane.  Equation (\ref{eq:RGfull2}) can be set to zero and solved to find the fixed point values of $e^2$.  This gives us both charged and uncharged fixed points, $e^2=0$ and $e^2=3 \epsilon/N$.  There are always four real fixed points with $e^2=0$, which are given by:

\begin{subequations}
\begin{align}
e^2&=0 \quad  u=0 \quad \quad \enskip \enskip \enskip \, \,  v=0 \label{eq:e0fixed1}\\
e^2&=0 \quad u=\frac{\epsilon}{N+4} \quad \enskip v=0 \label{eq:e0fixed2}\\
e^2&=0 \quad  u=0 \quad \quad \enskip \enskip \enskip \, \, v=\frac{\epsilon}{5} \label{eq:e0fixed3}\\
e^2&=0 \quad u=\frac{\epsilon}{5N-4} \quad v=\frac{(N-2)\epsilon}{5N-4}. \label{eq:e0fixed4}
\end{align}
\end{subequations}

These fixed points are identified (see \cite{chaikin2000:cmp}) as Gaussian (\ref{eq:e0fixed1}), Wilson-Fisher or Heisenberg (\ref{eq:e0fixed2}), Ising (\ref{eq:e0fixed3}), and cubic (\ref{eq:e0fixed4}) fixed points.  Note that the $z$ fields can be separated into their real and imaginary parts, and when $v=0$ the same fixed point structure appears as in the $O(2N)$ field theory with quartic interaction $(\sum^{2N}_{\alpha} \phi_{\alpha} \phi_{\alpha})^2$.  The addition of $v$ breaks the $O(2N)$ symmetry of the model, since it contains the interaction $\sum^{2N}_{\alpha} \phi^4_{\alpha}$, which is typically referred to as cubic anisotropy.  The other fixed point with $v \neq 0$, $u=0$ is referred to as the Ising fixed point, since it corresponds to $2N$ independent copies of the Ising field theory.

It is well known that in the $O(n)$ model (to first order in $\epsilon$) the Wilson-Fisher fixed point is stable for $n<4$, and for $n>4$ the cubic fixed point becomes the only stable one.  In our case $2N=n$, and we indeed observe exactly this behavior in our flow equations below and above $N=2$.

We now go on to discuss the fixed points when $e^2=3 \epsilon/N$.  The two fixed points with $v=0$ are given by:

\begin{equation}
\label{eq:v0fixed}
\begin{split}
e^2&=3 \epsilon/N \\  
u_{\pm}&=\frac{\epsilon}{2N(N+4)} \left( N+18 \pm \sqrt{(N+18)^2 - 216(N+4)} \right) \\
v&=0.
\end{split}
\end{equation}

\begin{figure}[!t]
\centerline{\includegraphics[angle=0,width=0.7\columnwidth]{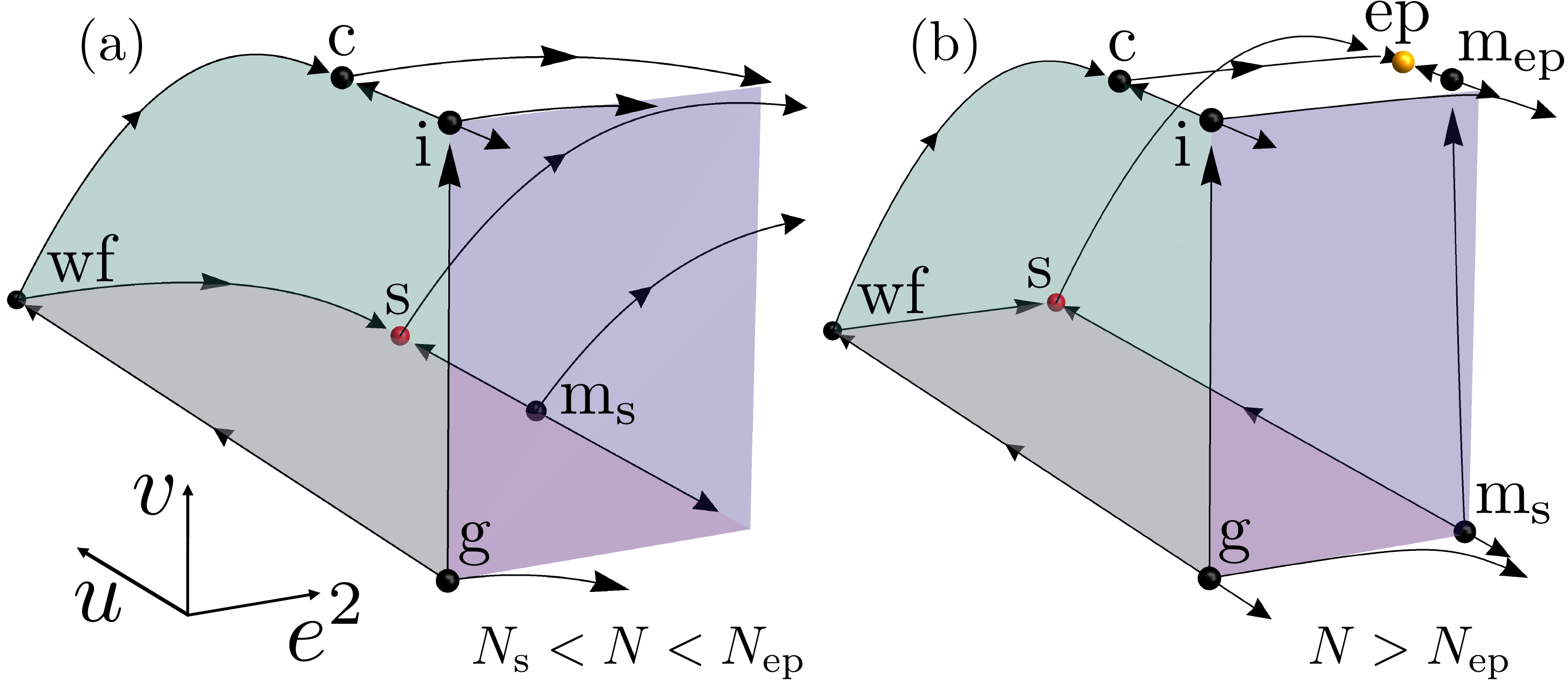}}
\caption{Here we show our flow diagrams for $N_{\mathrm{s}}<N<N_{\mathrm{ep}}$ and $N>N_{ep}$, this time with all of the fixed points marked.  The labels correspond to Gaussian (g), Wilson-Fisher (wf), Ising (i), cubic (c), symmetric deconfined (s), symmetric multicritical ($\mathrm{m}_{\mathrm{s}}$), easy-plane deconfined (ep) and easy-plane multicritical ($\mathrm{m}_{\mathrm{ep}}$).  ep is the only fixed point with all directions irrelevant in the $r=0$ plane.}
\label{fig:rgflowssupp}
\end{figure}

These two fixed points become real when $N>182.9516 = N_{\mathrm{s}}$.  Of these, the fixed point at $u_{+}$ is the symmetric deconfined fixed point, which is stable in the $r=0, v=0$ plane.  The multicritical point at $u_{-}$ has one relevant direction in the $r=0,v=0$ plane.  We note that when $v \neq 0$, this causes a runaway flow from the deconfined fixed point, which shows up as a first order transition in our lattice simulations.  For this value of $N$ (corresponding to panel (a) of Fig. \ref{fig:rgflowssupp}) there are no stable fixed points in the $r=0$ plane.  

We now move to the final fixed points that appear as $N$ is increased even further, which is the main result of this paper.  This corresponds to the finite solution for $v$ in Eqn (\ref{eq:RGfull4})

\begin{equation}
\label{eq:fixedep}
\begin{split}
e^2&=3 \epsilon/N \\  
u_{\pm}&=\frac{\epsilon}{2N(5N-4)} \left( N+18 \pm \sqrt{(N+18)^2 - 1080(5N-4)} \right) \\
v_{\pm}&=\frac{1}{5} \left[ \epsilon \left( 1+ 18/N \right) -6u_{\pm} \right]
\end{split}
\end{equation}

Here the solutions for $u$ (and hence the solutions for $v$) become real when $N>5363.1341=N_{\mathrm{ep}}$.  The solution at $u_{+}$ is the easy-plane deconfined fixed point, which has all directions irrelevant in the $r=0$ plane and is hence stable.  The other solution at $u_{-}$ is a multicritical point denoted by $m_{\mathrm{ep}}$ in Fig. \ref{fig:rgflowssupp}.  The easy-plane deconfined fixed point describes the criticality observed in our lattice model at large $N$. 

We note at this point that the values of $N_{s}$ and $N_{ep}$ obtained from the $\epsilon$ expansion at first order are notoriously unreliable.  The value of $N_{s}$ is believed to extend all the way down to $N=2$ based on the observed deconfined criticality in spin-$1/2$ systems \cite{kaul2013:qmc}. Likewise we observe $N_{\mathrm{ep}} \approx 20$ from our lattice simulations.  It is interesting to note that the $\epsilon$ expansion gives $N_{\mathrm{ep}}$ that is an order of magnitude larger than $N_{\mathrm{s}}$, which is observed in numerical simulations.

\end{document}